\documentclass[prl,twocolumn,amssymb,amsmath]{revtex4-1}
\usepackage{graphicx}
\usepackage{epsfig,color}
\usepackage{amsmath,amssymb}
\begin{document}

\title{Simulating 2+1d Lattice QED with dynamical matter using ultracold atoms}

\date{\today}

\author{Erez Zohar$^1$, J. Ignacio Cirac$^2$, Benni Reznik$^1$}

\begin{abstract}
\begin{center}
$^1$School of Physics and Astronomy, Raymond and Beverly Sackler
Faculty of Exact Sciences, Tel Aviv University, Tel-Aviv 69978, Israel.

$^2$Max-Planck-Institut f\"ur Quantenoptik, Hans-Kopfermann-Stra\ss e 1, 85748 Garching, Germany.
\end{center}
We suggest a method to simulate lattice compact Quantum Electrodynamics (cQED) using ultracold atoms in optical lattices, which includes dynamical Dirac fermions in 2+1 dimensions. This allows to test dynamical effects of confinement as well as 2d flux loops deformations and breaking, and to observe Wilson-loop area-law.
\end{abstract}

\maketitle

Recent progress in the manipulation and control of atomic systems has
boosted the interest in the emerging field of quantum simulation
 \cite{Blatt2012,Bloch2012,Lewenstein2012}.
So far, most of the theoretical and experimental effort has concentrated
in simulating condensed matter systems. The field of quantum simulation,
however, may also have a strong impact in high energy
physics (HEP) \cite{CiracZoller}, lending us the
possibility of observing intriguing phenomena emerging from the standard
model, or even allowing us to answer some questions which cannot
be addressed with standard techniques of lattice QCD. In contrast to its
condensed matter physics counterpart, the field of quantum simulation of
HEP models is almost unexplored, and presents new theoretical and
experimental challenges.

In the standard model of HEP, forces
are mediated through gauge bosons, which are the excitations of gauge
fields. Thus, gauge fields, either abelian or non-abelian, play a central
role in particle physics. As such, gauge theories have been a subject of
continuous research over the last decades, theoretically, numerically and
experimentally.  One important property of such theories is known as
confinement \cite{Wilson,KogutSusskind,Polyakov,BanksMyersonKogut},
which is manifested in real-world QCD, in the structure of hadrons: Free quarks cannot be found in nature, due to the phenomenon of confinement, which "holds them together" forming hadrons.
Confinement is known to hold in non-abelian gauge theories (not only QCD), but it is also
manifested in abelian theories, such as U(1) - QED: in the lattice case -
Compact QED (cQED). There, in 3+1d, a phase transition takes place between
a confining phase (for strong couplings) and a Coulomb phase (for weak
ones), and in 2+1d confinement takes place for all values of the coupling
constant \cite{Polyakov,BanksMyersonKogut,DrellQuinnSvetitskyWeinstein,KogutLattice,BenMenahem}.

Recently, some suggestions for simulations of QFT and HEP models have been proposed.
This includes the simulations of dynamic scalar \cite{Retzker2005} (vaccum entanglement) and fermionic fields \cite{Cirac2010} (the interacting Thirring and Gross-Neveu models).
Simulation proposals for fermions in Lattice QFT, either free or in external non-dynamical gauge fields include Axions and Wilson fermions \cite{Bermudez2010}, Dirac fermions in curved spacetime \cite{Boada2011} and general quantum simulators of QFTs and topological insulators \cite{Mazza2012}.
On the other hand, some simulations were proposed for dynamic gauge fields but with no fermions, using BECs \cite{Zohar2011} or single atoms \cite{SpinGauge} in optical lattices, where the first simulates the abelian Kogut-Susskind Hamiltonian \cite{KogutLattice} and the latter - a truncated ("Spin-Gauge") version of it (other examples of truncated models are \cite{Horn1981,Orland1990,Wiese1997}).
More simulations of pure-gauge $U(1)$ theories with ultracold atoms have been recently suggested \cite{Szirmai2011,Celi2012}.

The next step is the inclusion of dynamical matter (fermions) in the model,
allowing for simulation of full cQED. This is of special interest, both
for the possibility of probing confinement in a dynamic matter theory, as
well as for the exploration of problems which are
not amenable of a numerical description, due to the well-known Grassmanian integration sign problem \cite{Troyer2005}. The first proposals for the
simulation of a fermionic system with a dynamic gauge field \cite{Cirac2010,Banerjee2012} are, so far, restricted to 1+1d, where the emerging physical phenomena are limited due to the absence of magnetic fields and of multiple paths
connecting two points (see, however,
the QED (non-lattice) work \cite{Kapit2011}).

In this paper, we propose a spin-gauge model containing dynamic fermions,
that allows for simulation of nontrivial gauge field dynamics for 2+1
cQED, including 2d spatial effects such as flux loops. We show how to
construct a gauge invariant model of a U(1) gauge field coupled to 'naive'
fermions, discuss how to create various initial states, and suggest
several possible
experiments, such as the observation of broken flux lines and
manifestation of Wilson's area-law of confinement \cite{Wilson}, realizing the gedanken
experiment proposed in \cite{Topological}.

\emph{The Simulating System}. Consider a 2d spatial optical lattice \cite{Lewenstein2012}, filled with fermionic and bosonic atoms. Each vertex $\mathbf{n}=\left(n_1,n_2\right)$ is occupied by a single fermion, either of species $C$ or $D$, described by the local Hilbert spaces spanned by eigenstates of the fermionic number operators $N_{\mathbf{n}}^{C,D}$ (with the annihilation operators
 $c_{\mathbf{n}}$, $d_{\mathbf{n}}$). The two species form a spinor at each vertex:
\begin{equation}
\psi_{\mathbf{n}}=\left(\begin{array}{c}
c_{\mathbf{n}}\\
d_{\mathbf{n}}
\end{array}\right)
\end{equation}
We define the local charge \footnote{Note that this definition of charges is similar to the charge in the second quantized Dirac
field. There, for the lower entry of spinor, we swap the fermionic
creation and annihilation operators, and thus get that this charge
transforms to the $N_C-N_D$.},
\begin{equation}
Q_{\mathbf{n}}=\psi_{\mathbf{n}}^{\dagger}\psi_{\mathbf{n}}-1=N_{\mathbf{n}}^{C}+N_{\mathbf{n}}^{D}-1
\end{equation}
  Each link (in directions $k=1,2$) is occupied by a single boson, of one of $2l+1$ species, forming an angular momentum multiplet, described by eigenstates of $L_{z,\mathbf{n}}^k$ (See also the figure in \cite{Supp}).

  We call this system "the primitive theory" since the gauge-invariant effective theory will arise from it, by adding a constraint, as will be shown later, but first we shall describe the (separate) primitive theories for the bosons and the fermions. Similar fermionic theories can be found, for example, in \cite{Bermudez2010,Boada2011,Mazza2012}.

The bosons primitive dynamics is governed by the Hamiltonian
\begin{equation}
H_{p}^{b}= \underset{\mathbf{n},k}{\sum} \left( \mu \left( L_{z,\mathbf{n}}^k \right)^2 + 2\beta L_{x,\mathbf{n}}^k \right)
 + \Omega \underset{\left\langle i,j \right\rangle}{\sum} \left( L_{x,i} L_{x,j} + L_{y,i} L_{y,j} \right)
 \label{Hpb1}
\end{equation}
where the second sum is on nearest-neighbor (intersecting) links.

The fermions primitive dynamics is governed by the lattice Dirac Hamiltonian
\begin{equation}
H_{p}^{f}= i\eta\underset{\mathbf{n},k}{\sum}\left(\psi_{\mathbf{n}}^{\dagger}\sigma_{k}\psi_{\mathbf{n+\hat k}}-H.C.\right) + M\underset{\mathbf{n}}{\sum}\psi_{\mathbf{n}}^{\dagger}\sigma_{z}\psi_{\mathbf{n}}
\label{Hpf1}
\end{equation}

\emph{Imposing Gauge invariance}. We wish to impose gauge invariance to the system. In order to do so, we constrain the generators of gauge symmetry $G_\mathbf{n}$ to zero, by adding the
"Gauss's Hamiltonian" $H_G$ to the two primitive Hamiltonians. This Hamiltonian involves an interaction between the two primitive systems.
The generators of local gauge transformations are
\begin{equation}
G_{\mathbf{n}}=L_{z,\mathbf{n}}^{1}+L_{z,\mathbf{n}}^{2}+L_{z,\mathbf{n - \hat 1}}^{1}+L_{z,\mathbf{n - \hat 2}}^{2}-\left(-1\right)^{n_1+n_2}Q_{\mathbf{n}}
\end{equation}
and the constraint is implemented by the Gauss Hamiltonian,
$H_{G}=\lambda\underset{\mathbf{n}}{\sum}G_{\mathbf{n}}^{2}$
and thus $\lambda$ must be the highest energy scale: $\lambda \gg \mu , \beta , \Omega , \eta , M$.

As $\lambda$ is the largest energy scale, we wish to derive
an effective theory for $H_G$'s ground sector \cite{Soliverez1969}, which will introduce gauge invariance to the system.
In the first order, we get the bosonic "electric Hamiltonian"
$H_{E} = \mu \underset{\mathbf{n},k}{\sum} \left( L_{z,\mathbf{n}}^k \right)^2$,
and the fermionic "Mass Hamiltonian"
$H_{M} = M\underset{\mathbf{n}}{\sum}\psi_{\mathbf{n}}^{\dagger}\sigma_{z}\psi_{\mathbf{n}}$

In the second order we get four contributions:
(1) The bosonic "magnetic" plaquette terms:
$H_{B} = - \frac{2 \Omega ^2}{\lambda} \underset{\mathbf{n}}\sum \left( L_{+,\mathbf{n}}^1 L_{-,\mathbf{n+ \hat 1}}^2 L_{+,\mathbf{n+ \hat 2 }}^1 L_{-,\mathbf{n}}^2 + H.C. \right)$
as well as the bosonic $H'_B$, which has no equivalent in cQED, but is gauge-invariant and contributes no interesting dynamics (see \cite{SpinGauge} for details).
(2) The "minimal coupling" Dirac terms:
$H_D = -\frac{i\beta\eta}{\lambda}\underset{\mathbf{n},k}{\sum}\left(\psi_{\mathbf{n}}^{\dagger}\sigma_{k}\psi_{\mathbf{n+ \hat k}}L_{s,\mathbf{n}}^{k}-H.C.\right)$
(where $s=(-)^{n_1+n_2}$).
The other two contributions are less important, and are described in the supplementary material \cite{Supp}.

Next we make the sign change $L_{z,\mathbf{n}}^{k}\longrightarrow\left(-1\right)^{\mathbf{n}}L_{z,\mathbf{n}}^{k}$. Besides inverting the sign of $L_z$ on odd links, we also swap the $L_{\pm}$ operators there (leaving $L_x$ invariant). That results, as in the pure spin-gauge theory, with correct signs in the plaquette term
\begin{equation}
H_{B} = - \frac{2 \Omega ^2}{\lambda} \underset{\mathbf{n}}\sum \left( L_{+,\mathbf{n}}^1 L_{+,\mathbf{n+ \hat 1}}^2 L_{-,\mathbf{n+ \hat 2 }}^1 L_{-,\mathbf{n}}^2 + H.C. \right)
\end{equation}
and the gauge generators
\begin{equation}
G_{\mathbf{n}}=(-1)^{n_1+n_2}\left(L_{z,\mathbf{n}}^{1}+L_{z,\mathbf{n}}^{2}-L_{z,\mathbf{n - \hat 1}}^{1}-L_{z,\mathbf{n - \hat 2}}^{2}-Q_{\mathbf{n}}\right)
\end{equation}
This also gives rise to the correct (minimally coupled) Dirac Hamiltonian
\begin{equation}
H_{D}=\frac{i\eta\beta}{\lambda}\underset{\mathbf{n},k}{\sum}\left( \psi_{\mathbf{n+\hat k}}^{\dagger} \sigma_{k} \psi_{\mathbf{n}}L_{-,\mathbf{n}}^{k}-H.C.\right)
\end{equation}
where we identify the Dirac matrices $\mathbf{\alpha}_{1,2}$ as the
Pauli matrices $\mathbf{\sigma}_{1,2}$ (The Dirac $\beta$ matrix is $\sigma_{z})$.

Thus, we have obtained the effective Hamiltonian (neglecting $H'_B$):
\begin{equation}
H_{l,cQED} = \alpha^{-1}\left(H_{E}+H_{B}+H_{D}+H_{M}\right)
\label{Heff}
\end{equation}
where $\alpha$ is a rescaling factor: $\alpha=\frac{2}{g^{2}}\mu=\frac{4 l^2\left(l+1\right)^2\Omega^{2}g^{2}}{\lambda}=\frac{2\eta\beta \sqrt{l\left(l+1\right)}}{\lambda}=\frac{M}{m}$, introducing the gauge theory coupling constant $g$ and fermions mass $m$ \cite{Supp}. Equation (\ref{Heff}) is the Kogut-Susskind Hamiltonian of lattice cQED with dynamical Dirac Fermions.

We wish to emphasize that the previous effective Hamiltonian calculation does not depend on $l$, and thus the primitive theory and the constraint are achieved for other values of $l$, the same effective Hamiltonian results.

\emph{Realization of the model.} In order to realize the model for $l=1$, for the intra-bosonic terms one can use the methods presented in \cite{SpinGauge}, which generalize \cite{Mazza2010}. Almost all the required intra-bosonic terms in (\ref{Hpb1}) (including the ones in $H_G$, but not the linear $L_x$ terms) are obtained if one sets, in \cite{SpinGauge} terms, $q_{\mathbf{n}}=0$ everywhere. To get the linear $L_x$ terms of (\ref{Hpb1}), one must add an on-site species rotation term which can be achieved using Raman lasers, since $L_{+}=\underset{m=-l}{\overset{l-1}{\sum}}\sqrt{l\left(l+1\right) - m\left(m+1\right)}a_{m+1}^{\dagger}a_{m}$ (where $a_{m}$ is the local $m$th bosonic species annihilation operator). This does not affect the effective calculation since these are local terms.

For details on the fermionic interactions, see \cite{Supp}. The main issue is now the realization of the Boson-Fermion interaction terms in $H_G$. The required terms, for each vertex $\mathbf{n}$, are of the form
\begin{equation}
\lambda\xi\left(\mathbf{n}\right) \left(L_{z,\mathbf{n}}^{1}+L_{z,\mathbf{n}}^{2}+L_{z,\mathbf{n - \hat 1}}^{1}+L_{z,\mathbf{n - \hat 2}}^{2}\right) \left(N^C_{\mathbf{n}} + N^D_{\mathbf{n}}\right)
\label{reqint}
\end{equation}
where $\xi\left(\mathbf{n}\right) = -2 \left(-1\right)^{n_1+n_2}$
For a start, we choose the atomic hyperfine levels of the fermionic degrees of freedom:
C and D fermions belong to the hyperfine levels
$\left|F_{C}=\frac{1}{2},m_{C}\right\rangle $ ,
$\left|F_{D}=\frac{3}{2},m_{D}\right\rangle $  respectively, where $m_{C},m_{D}$
are constant on each vertex: The hyperfine levels are arranged such that the other values of $m$, on each
hyperfine manifolds, are of a much larger energy, and thus are not
accessible if they are not initially populated.

The (\ref{reqint}) terms are derived as scattering interactions between
two particles: bosons with $F_{b}=1$ and each of the fermionic species.
The first quantized interaction Hamiltonian corresponding to such
scattering processes is $H_{sc}=\underset{F}{\sum}g_{F}P_{F}$,
where the summation is on the total angular momentum of the two scattered
particles, $\{g_{F}\}$ depend on the S-wave scattering length, and
thus are tunable using optical Feshbach Resonances \cite{Fedichev1996,Bohn1997,Fatemi2000}, and $P_{F}$ are projection
operators to the subspaces of different total angular momenta. These operators
can be built using the different values of $\mathbf{F}_{C}\cdot\mathbf{F}_{b}$, $\mathbf{F}_{D}\cdot\mathbf{F}_{b}$
for each value of total $\mathbf{F}$, and thus one can express the scattering Hamiltonians for $C,D$ as
\begin{multline}
H_{sc}^{C}=\widetilde C_0 + \widetilde C_1 \mathbf{F}_{C}\cdot\mathbf{F}_{b} \\
H_{sc}^{D}=\widetilde D_{0}  + \widetilde D_{1}  \mathbf{F}_{D}\cdot\mathbf{F}_{b} + \widetilde D_{2} \left(\mathbf{F}_{D}\cdot\mathbf{F}_{b}\right)^{2}
\end{multline}
where $\widetilde C_i = \widetilde C_i \left(g_{\frac{1}{2}},g_{\frac{3}{2}}\right) , \widetilde D_i =  \widetilde D_i \left(g_{\frac{1}{2}},g_{\frac{3}{2}},g_{\frac{5}{2}}\right)$

These Hamiltonians are next plugged into the second quantized interaction terms $\int d^{3}x\Psi_{i}^{C\dagger}\Phi_{j}^{\dagger}H_{sc}^{C}\Psi_{k}^{C}\Phi_{l}$,
where $\Psi_{k}^{C},\Phi_{l}$ are the fermionic and bosonic second
quantized wavefunctions. Due to the energy spectrum of the fermions,
the only possible processes are such with $i=k=m_{C}$. Thus no angular
momentum transfer can take place for the bosons as well, and hence
$j=l=m_{b}$, and only the $z$ components of the angular momentum
vectors contribute. We finally get, on each vertex,
\begin{multline}
\underset{m_{b}}{\sum}\left(C_{0}+C_{1}m_{C}m_{b}\right)c^{\dagger}_\mathbf{n} c_\mathbf{n} a_{m_{b}}^{\dagger}a_{m_{b}} + \\
\underset{m_{b}}{\sum}\left(D_{0}+D_{1}m_{D}m_{b}+D_{2}f\left(m_{D},m_{b}\right)\right)d^{\dagger}_\mathbf{n}d_\mathbf{n}a_{m_{b}}^{\dagger}a_{m_{b}}
\end{multline}
where the bosonic operators correspond to each of the neighboring links, $C_i,D_i$ are products of the coefficients $\widetilde C_i,\widetilde D_i$ and the appropriate overlap integrals,
and $f\left(m_{D},m_{b}\right)$ is quadratic.

In order to get (\ref{reqint}), we impose conditions on the scattering coefficients $g_F$ (controlled by Feshbach resonances):
$D_{2}\left(g_{\frac{1}{2}},g_{\frac{3}{2}},g_{\frac{5}{2}}\right)=0$,
$D_{0}\left(g_{\frac{1}{2}},g_{\frac{3}{2}},g_{\frac{5}{2}}\right)=C_{0}\left(g_{\frac{1}{2}},g_{\frac{3}{2}}\right)$
and
$D_{1}\left(g_{\frac{1}{2}},g_{\frac{3}{2}},g_{\frac{5}{2}}\right)\left|m_{D}\right|=C_{1}\left(g_{\frac{1}{2}},g_{\frac{3}{2}}\right)\left|m_{C}\right| = -2 \lambda$
the first one eliminates the quadratic terms. The second one results in
terms of the form $C_{0}\left(N_{\mathbf{n}}^{C}+N_{\mathbf{n}}^{D}\right)N^{b}$. However, the bosonic effective Hamiltonian sets that
$N^{b}=1$ everywhere around the bosonic lattice, and thus this term is merely the total number of fermions in the system, which is an ignorable constant in the Hamiltonian.
Setting that at the vertex $\mathbf{n}$, $m_{C} = \left(-1\right)^{n_1+n_2}\left|m_{C}\right|, m_{D} = \left(-1\right)^{n_1+n_2}\left|m_{D}\right|$, we are only left with one type of term, which is, due to the third condition,
$-2 \lambda \left(-1\right)^{n_1+n_2} \left(N_{\mathbf{n}}^{C}+N_{\mathbf{n}}^{D}\right)L_{z}$, which is the requested interaction.

\emph{No Interactions vacuum and excited states.}
Suppose first that the bosonic and fermionic interactions ($H_B , H_D$) are turned off - i.e., the system is
pure-gauge (the charges are static and uncoupled) and in the strong coupling limit. Hence the gauge field vacuum does not
contain any excited links: all the bosonic atoms are in their $m=0$ state. This is the exact QED vacuum in the extreme strong limit, since no plaquette terms contribute.

As for the fermionic content, each vertex can contain four different fermionic contents. The state with no charges corresponds to
the "Dirac Sea" - i.e., all the vertices are filled with $D$ atoms. In such a state, the local "mass" is $-M$. Since energies are
measured above the Dirac sea, we would like to measure these "masses" above $-M$, i.e. shift the energies with the product of $M$
with the total number of vertices. This is the vacuum in case of no charges. The three excited states correspond to different simulated fermionic contents, as
summarized in the following table; Our "particle" is $q$, with a positive charge of $+1$, and its "anti-particle" is $\bar q$, with a negative charge of $-1$. Note that the masses are relative to $-M$:

    \begin{tabular}{ | l | l | l | p{3cm} |}
    \hline
    Real Content & "Mass" & "Charge" & Simulated Content \\ \hline
    $\left| 0 \right\rangle_C \left| 1 \right\rangle_D$ & $0$ & $0$ & Empty vertex \\ \hline
    $\left| 0 \right\rangle_C \left| 0 \right\rangle_D$ & $M$ & $-1$ & $\bar q$ \\ \hline
    $\left| 1 \right\rangle_C \left| 1 \right\rangle_D$ & $M$ & $+1$ & $q$ \\ \hline
    $\left| 1 \right\rangle_C \left| 0 \right\rangle_D$ & $2M$ & $0$ & $q \bar q$ \\ \hline
    \end{tabular}

\emph{Initial state preparation.}
The system is initially prepared in the gauge+fermions ground state, which is an eigenstate of $H_G + H_E + H_M$ (Bosons - $m=0$ everywhere, and all the vertices are occupied only with $D$ fermions). Then, these three Hamiltonians
can be turned on, without changing the state but imposing the constraint.

There are several interesting initial states one can obtain, even before turning $H_D$ on. These include:
(a) Obtaining the QED static vacuum (up to first order, if one works with $l=1$), by increasing $\Omega$ adiabatically, keeping $\frac{2\Omega^2 l^2 \left(l+1\right)^2}{\lambda} \ll \mu$;
(b) Obtaining a "loop sea" by increasing $\Omega$ adiabatically again, but with $\frac{2\Omega^2 l^2 \left(l+1\right)^2}{\lambda} \gg \mu$, dressing to a state with a lot of loops, but with larger amplitudes, and then, by lowering the value of $\Omega$, one can obtain strong regime QED dynamics again;
(c) Creating, using single-addressing lasers \cite{Bakr2009,Weitenberg2011}, zeroth order excitations: one can create large closed flux-loops as in the pure gauge case, but also and/or place $C$ charges and empty vertices to form zeroth-order flux tubes. Afterwards $H_B$ can be turned on adiabatically to obtain QED dynamics.

\begin{figure}[t]
\includegraphics[scale=0.3]{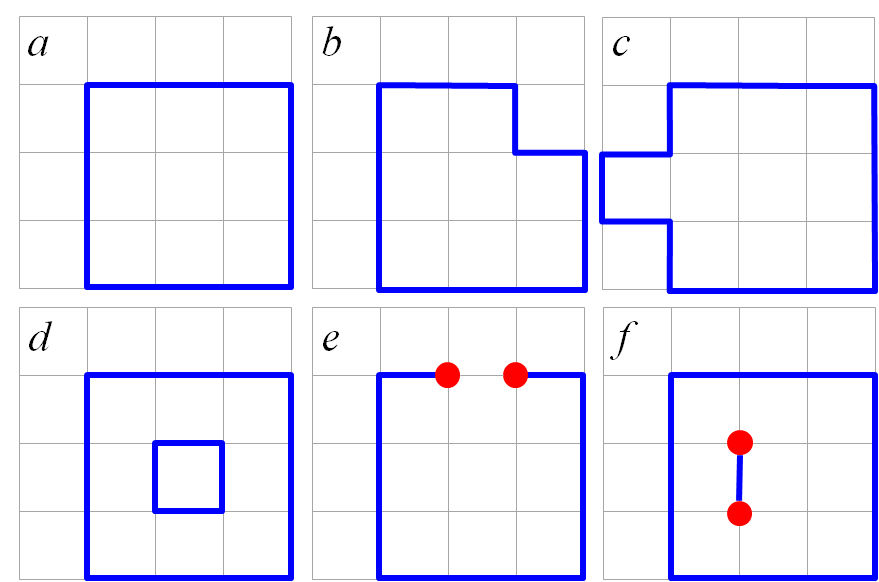}
\caption{Examples of gauge field and matter dynamics, as explained in the text, for an initial state (using single addressing lasers) of a single large flux loop (a). (b-f) show possible outcomes of first order processes:
(b-d) result from single operations of $H_B$: (b) loop decreasing, (c) loop increasing, (d) exciting a non-connected plaquette; (e-f) result from single operations of $H_D$: (e) flux loop breaking and creation of two charges, (f) creation of a meson outside the loop with a new flux tube connecting them.}
\label{fig2}
\end{figure}

\emph{Inclusion of Matter Dynamics.}
Given any of the initial states (a),(b) or (c), the field-matter interaction $H_D$ can be turned on (in order to remain in the strong limit, the parameters must obey $\frac{2\Omega^2 l^2 \left(l+1\right)^2}{\lambda} \ll \frac{\eta\beta \sqrt{l\left(l+1\right)}}{\lambda} \ll \mu$ \cite{Supp}).
Then, an interplay between the fermion masses and the fermionic interaction terms introduces changes in the flux
loops/tubes structure, if one waits long enough.
For example, suppose $2M=\mu L$, for some integer $L$. Then, if there is a flux tube/loop longer than $L$, we expect it to break
to two flux tubes, generating two new fermions, transferring the energy of the "disappearing links" to the masses of the
new fermions. On the other hand, if the fermion mass is small (or zero) we expect the flux tube to break and form new fermions constantly.

Another possibility is to move slowly (externally) one of the charges of an initially prepared flux-tube and
stretching it, which should lead to it breaking apart.

Examples of matter dynamic processes are shown in figure (\ref{fig2}).

\emph{Wilson's Area law observation.}
Another possible measurement which can be done in this system is a test of Wilson-loop area law, as a probe for confinement.
This can be done by an interference of two mesons in superposition. This cannot be done in real-world QED, where it is only a gedanken experiment \cite{Topological}. Here we propose to realize it in the quantum simulation.
For that, we need the fermions to be static and their mass is unnecessary, and hence $H_{M},H_{f}$
are turned off (and thus, effectively, $H_D$ as well). In fact, for a proof of principle of this experiment, $H_B$ is not needed too.

We start, in the extreme strong limit, with a state of a single flux-tube with
length $R$ over the Dirac Sea -- $\left|R\right\rangle $, emanating
between the vertices $\left(m,n\right)$ and $\left(m+R,n\right)$,
such that $Q_{m,n}=1,Q_{m+R,n}=-1$. This state is just a meson in the strong coupling limit, and it is an eigenstate of the Hamiltonian.
Then, as in \cite{Topological}, we wish create a superposition
of two states, $\left|R\right\rangle $ and $\left|R+1\right\rangle $: i.e., to transfer the $C$ fermion, with a probability $\frac{1}{2}$, from $(m,n)$ to $(m-1,n)$. We shall treat this subspace as a two level system,
denoted by $\left|\downarrow\right\rangle =\left|R\right\rangle $,$\left|\uparrow\right\rangle=\left|R+1\right\rangle $, with the Hamiltonian $H_{0} = \mu\left|\uparrow\right\rangle \left\langle \uparrow\right|$.

The required superposition is generated as in Ramsey interferometry \cite{Supp}: we apply on the initial state
$\left| \downarrow \right\rangle$
the unitary operation $U = e^{-i \frac{\pi}{4} \sigma_y}$,
and get the state $U\left| \downarrow \right\rangle = - \left| \downarrow_x \right\rangle$.
Next, wait a long time $T \gg \tau$, during which the state evolves to
$\frac{1}{\sqrt{2}}e^{-i\mu RT}\left(\left|\downarrow\right\rangle -e^{-i\mu T}\left|\uparrow\right\rangle \right)$. Operating on the state with $U$ again yields
$\frac{1}{\sqrt{2}}\left(\left|\downarrow_{x}\right\rangle + e^{-i\mu T}\left|\uparrow_{x}\right\rangle \right)$ (neglecting the global phase).
Switching back to the $\sigma_{z}$ basis, recalling the definitions of our effective two-level system, we get
that the probability to find a fermion C in the vertices $\left(m,n\right),\left(m-1,n\right)$
are
\begin{equation}
P_{m,n}=\sin^{2}\left(\frac{\mu LT}{2}\right) ; P_{m-1,n}=\cos^{2}\left(\frac{\mu LT}{2}\right)
\end{equation}
since $A=LT$ (above $L=1$), this manifests the area law in a confining phase, and thus,
within this system a realization of the gedanken experiment proposed there is possible.
In future generalizations, this method may serve as phase transition probe.

\begin{figure}
\includegraphics[scale=0.3]{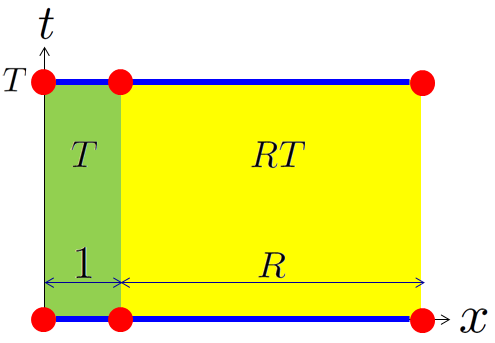}
\caption{Description of the area-law experiment in space-time, as described in the text. The interference phase depends on the green area.}
\end{figure}

\emph{Summary.} In this paper, we have proposed a method to add dynamic fermions to the Spin-Gauge model introduced in \cite{SpinGauge}. This allows to simulate 2+1d cQED using ultracold atoms in optical lattices, enabling the observation
 of confinement in a dynamic matter theory, and suggesting a detour to the well known Grassmanian integration sign problem, encountered in Monte-Carlo simulations \cite{Troyer2005}. Simulations of our model on a small lattice may be used to check the effects of imperfections of the parameters on the dynamics. We have suggested several possible measurements, which can show the emergence of dynamic charges, the breaking of flux-tubes and the area law behavior of the confining potential.

\emph{Acknowledgements.}
BR acknowledges the support of the Israel Science Foundation, the German-Israeli
Foundation, and the European Commission (PICC). IC is partially supported by the EU project
AQUTE.

\bibliography{ref}

\section{Supplemental material}

\subsection{Tailoring the fermion interactions}
\begin{figure}[h]
\includegraphics[scale=0.5]{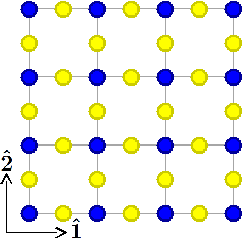}
\caption{The simulating lattice. Every link is occupied by a single boson (yellow circle) and every vertex by either zero, one or two fermions (blue circles).}
\label{fig1}
\end{figure}

The intra-fermionic terms (in the fermionic primitive Hamiltonian) can be obtained if one tailors the fermionic Raman transitions and chemical potentials properly. Note that the required hopping terms, achieved using Raman lasers, include "species rotation", i.e.: for example, when a $C$ fermion hops to a neighboring vertex it becomes a $D$ - this is manifested by the Pauli matrices $\sigma_x,\sigma_y$ in the interactions. However, this rotation can be avoided, if one initially defines slightly different primitive theories as we shall next describe. Other required intra-fermionic terms are the ones on each vertex, resulting from $\lambda Q_{\mathbf{n}}^2$ in $H_G$: some add up to a constant - the total number of fermions in the system, and can be disregarded, and the others are local fermion scattering terms:
$2\lambda  N_{\mathbf{n}}^C N_{\mathbf{n}}^D$.

Now we describe an alternative method to get the same effective theory as in the paper, but with same-species fermionic tunneling and slightly different primitive Hamiltonians.
We start by defining \emph{another} spinor at each vertex, rather than $\psi_{\mathbf{n}}$,
\begin{equation}
\chi_{\mathbf{n}}=\left(\begin{array}{c}
c_{\mathbf{n}}\\
d_{\mathbf{n}}
\end{array}\right)
\end{equation}
(note that the charges remains the same, as they are invariant for local fermionic species swapping.)
and the primitive Hamiltonians are
\begin{widetext}

\begin{equation}
H_{p}^{b}= \mu\underset{\mathbf{n},k}{\sum}  \left( L_{z,\mathbf{n}}^k \right)^2 + 2\beta\underset{\mathbf{n}}{\sum} \left( L_{x,\mathbf{n}}^1 + \left(-1\right)^{n_1+n_2} L_{x,\mathbf{n}}^2 \right)
 + \Omega \underset{\left\langle i,j \right\rangle}{\sum} \left( L_{x,i} L_{x,j} + L_{y,i} L_{y,j} \right)
 \label{Hpb2}
\end{equation}
where the second sum in on nearest-neighbor (intersecting) links, and
\begin{equation}
H_{p}^{f}= i\eta\underset{\mathbf{n}}{\sum}\left(\chi_{\mathbf{n}}^{\dagger}\chi_{\mathbf{n+\hat 1}}-H.C.\right) +
\eta\underset{\mathbf{n}}{\sum}\left(\chi_{\mathbf{n}}^{\dagger}\sigma_{z}\chi_{\mathbf{n+\hat 2}}+H.C.\right) +
M\underset{\mathbf{n}}{\sum}\left(-1\right)^{n_1+n_2}\chi_{\mathbf{n}}^{\dagger}\sigma_{z}\chi_{\mathbf{n}}
\label{Hpf2}
\end{equation}
\end{widetext}
if one now performs the transformation
\begin{equation}
\chi_{\mathbf{n}}=\sigma_x^{n_1+n_2}\psi_{\mathbf{n}}
\end{equation}
and repeats the effective calculation for this model, the same effective Hamiltonian will result.

Since the charges remain the same, one could produce the bosonic-fermionic interactions exactly as described in the paper, where $\psi$ was the fundamental spinor, since they depend only on the invariant charge.

\subsection{More terms of the effective Hamiltonian}
In the paper we have described two types of second-order terms in the effective Hamiltonian. We hereby describe other two second-order contributions, which are irrelevant for our purposes:

\begin{enumerate}
\item Double operation of $H_{f}$ (fermionic tunneling):
Such terms may represent back-and-forth tunneling along a link. Some add up to the total number of fermions in the system,
which is an ignorable constant of motion, while others are C-D swap terms along a link. These terms are of order
$\frac{\eta^{2}}{\lambda}$ and thus can be attenuated (compared to the Dirac terms) if one sets
$\eta\ll\beta$.

\item Double operation of $H_{X}$: we
get, on every link, a term of the form
$-\frac{4\beta^{2}}{2\lambda}\left\{ L_{+},L_{-}\right\} =-\frac{4\beta^{2}}{\lambda}\left(\mathbf{L}^{2}-L_{z}^{2}\right)$.
the first term is a constant. The second is just a renormalization
of $\mu$.
\end{enumerate}

\subsection{Rescaling the Hamiltonian parameters}

In equation (9), we have defined the rescaling factor $\alpha$ which translates the experimental parameters into the simulated system parameters, introducing the coupling constant $g$ and the fermion mass $m$. The relation between the two families of parameters is given, as written in the main text, by
\begin{equation}
\alpha=\frac{2}{g^{2}}\mu=\frac{4 l^2\left(l+1\right)^2\Omega^{2}g^{2}}{\lambda}=\frac{2\eta\beta \sqrt{l\left(l+1\right)}}{\lambda}=\frac{M}{m}
\end{equation}
This relation contains the translations of the electric, magnetic, Dirac and mass terms.

One should note that the electric, magnetic and mass part contain constants of the theory - $g$ and $m$, and thus have some more freedom in their definitions. The Dirac kinetic part, however, contains, in the cQED Hamiltonian (setting the lattice spacing $a=1$) no physical constant, and thus the experimental parameters within it define $\alpha$. Thus, if we wish to examine the strong coupling limit, for example, demanding $g \gg 1$ from the electric part, we get that $\mu \gg \frac{\alpha}{2}$, which means
$\mu \gg \frac{\eta\beta \sqrt{l\left(l+1\right)}}{\lambda}$. On the other hand, from the magnetic part we get $\frac{4\Omega^2 l^2 \left(l+1\right)^2}{\lambda} \ll \alpha$, which corresponds to
$\frac{2\Omega^2 l^2 \left(l+1\right)^2}{\lambda} \ll \frac{\eta\beta \sqrt{l\left(l+1\right)}}{\lambda}$.
This yields the strong limit condition, in the presence of dynamic fermions, which is stated in the paper:
\begin{equation}
\frac{2\Omega^2 l^2 \left(l+1\right)^2}{\lambda} \ll \frac{\eta\beta \sqrt{l\left(l+1\right)}}{\lambda} \ll \mu
\end{equation}

One should also note that all this is
 true if one wishes to simulate "regular" cQED. However, if one chooses the parameters differently, the theory will still be gauge invariant (as long as the scale hierarchy is kept and the effective expansion is valid), but not the "regular" cQED theory.

\subsection{Comparison of simulating and simulated processes}

In terms of the simulating system, we shall denote the spinors as
\begin{equation}
\tilde \psi_{\mathbf{n}}=\left(\begin{array}{c}
q_{\mathbf{n}}\\
\bar q_{\mathbf{n}}^{\dagger}
\end{array}\right)
\end{equation}

Due to the Pauli matrices which couple nearest-neighbor vertices, it is only possible to either both create a particle and an anti-particle in the two ends of a link,
or to both annihilate such a pair. This is phrased in terms of the simulating system as the fact that creation of a $C$ fermion in a vertex means the annihilation of a
neighboring $D$, and vice-versa. Let us consider these equivalence of the simulating and simulated systems in some basic dynamic processes.

Fig. 4 shows some of these processes - considering the initial and final states in both terms, and the parts of the Hamiltonian which are responsible for them (only the specific relevant parts of the spinors are included).

\subsection{Creating the unitary transformation for the Ramsey interferometry}

Using Raman lasers, we introduce the interactions
\begin{equation}
\frac{i \eta}{\sqrt{2}} \left( c_{m,n}^{\dagger}c_{m-1,n} - c_{m-1,n}^{\dagger}c_{m,n} \right)
\end{equation}
 where $\eta \ll \lambda$, which are similar to the ones in $H_p^f$. These interactions violate the constraint, and thus we can use them in order to construct effective Hamiltonian terms
 $\frac{i \eta \beta}{\sqrt{2} \lambda} \left( c_{m,n}^{\dagger}c_{m-1,n}L_{(m-1,n)-} - c_{m-1,n}^{\dagger}c_{m,n}L_{(m-1,n)+} \right)$
 which can be rewritten as
\begin{equation}
H_y =   \frac{ \eta \beta}{\sqrt{2} \lambda} \sigma_y
 \end{equation}
 (Without loss of generality, we assume that $(m-1,n)$ is an even vertex). We turn these terms on and off suddenly, keeping them open a period of time $\tau = \frac{ \sqrt{2} \pi \lambda}{4 \eta \beta}$. $\tau \gg \lambda^{-1}$, which justifies the effective treatment. On the other hand, we choose the time $T$ (see the main text) to satisfy $\tau \ll T$, making $H_y$ strong and impulsive. Thus, the time evolution of the state when $H_y$ is on is equivalent to the unitary operation $U = e^{-i \frac{\pi}{4} \sigma_y}$, and if the initial state was $\left| \downarrow \right\rangle$, we now get $U\left| \downarrow \right\rangle = - \left| \downarrow_x \right\rangle$.

\begin{figure}[h!]
  \includegraphics[scale=0.9]{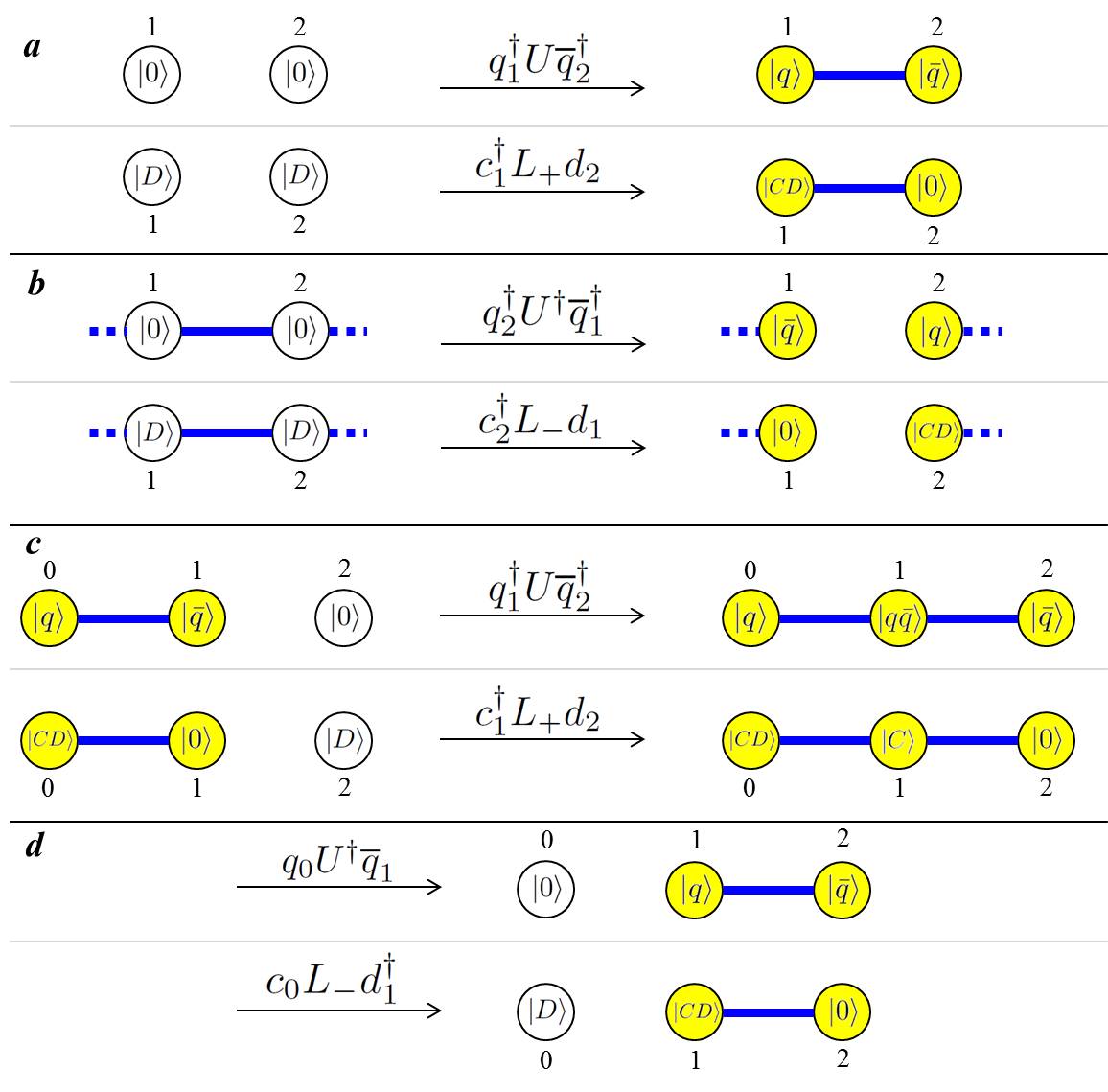}
  \caption{Possible fermionic processes, both in terms of the simulated system - QED (the upper rows) and of the simulating system (the lower rows), including the relevant operators. The initial states are on the left (a-c) and the final are on the right. (a) Creation of a particle-anti-particle flux tube (a meson). (b) Breaking a flux-tube. (c) "Stretching" a flux-tube. Notice that it is not really stretching, since a pair of particles appear in the middle. (d) Moving a meson: first process (c) occurs, and then a second stage, shown in (d), takes place. In this figure, in terms of the simulating system (lower rows),
  $\left| 0 \right\rangle_C \left| 0 \right\rangle_D = \left| 0 \right\rangle$,
  $\left| 1 \right\rangle_C \left| 0 \right\rangle_D = \left| C \right\rangle$,
  $\left| 0 \right\rangle_C \left| 1 \right\rangle_D = \left| D \right\rangle$,
  $\left| 1 \right\rangle_C \left| 1 \right\rangle_D = \left| CD \right\rangle$.
  }
\end{figure}

\end{document}